\documentclass{article}

\usepackage{graphicx}

\def\abstract#1{\vskip 7mm 
        \begin{center}{\large Abstract}\par \smallskip
                \begin{minipage}[c]{12cm}
                        \small #1
                \end{minipage}
        \end{center}
}
\def\title#1{\begin{center}{\Large\bf #1}\end{center}}
\def\author#1{\vskip 5mm \begin{center}{#1}\end{center}}
\def\address#1{\begin{center}{\it #1}\end{center}}
\def\beq{\begin{equation}}
\def\eeq{\end{equation}}
\def\beqa{\begin{eqnarray}}
\def\eeqa{\end{eqnarray}}
\makeatletter
\@ifundefined{lesssim}{}{}
\@ifundefined{gtrsim}{}{}
\def\vereq#1#2{\lower3pt\vbox{\baselineskip1.5pt \lineskip1.5pt
\ialign{$\m@th#1\hfill##\hfil$\crcr#2\crcr\sim\crcr}}}
\makeatother

\begin{document}

\title{%
  Inflation from String Theory
}
\author{%
  James M.\ Cline\footnote{E-mail:jcline@physics.mcgill.ca}
}
\address{%
 Department of Physics, McGill University, 3600 University St.,\\
Montr\'eal, Qc H3A 2T8 Canada
}

\abstract{I review some aspects of obtaining inflation from string theory, in
particular brane-antibrane inflation within the framework of KKLMMT, and
racetrack inflation.  Further, I discuss recent work on the problem of
reheating after brane-antibrane annihilation, and possible distinctive features
of production of cosmic string and brane defects at this time. }

\section{Introduction}

Many experiments are now yielding exciting, high-precision results about the
universe, concerning the cosmic microwave background,  large scale structure
through galaxy surveys, Lyman $\alpha$ forest observations and weak
gravitational lensing, and the dark energy via distant type Ia supernovae.
Particle physics experiments, on the other hand, are presently quite limited in
what they can tell us about physics at the most fundamental level,
for which string theory is a very attractive candidate theory. 
For these reasons many string theorists have turned to the early universe as
a possible laboratory for testing the theory.

In this talk I will focus on string theory as a source of inflation, various
aspects of which have been investigated in  references
\cite{realistic}-\cite{BC}.  I will start by reviewing
the KKLT \cite{KKLT} and KKLMMT \cite{KKLMMT} picture, then describe
our attempt to embed this in a more realistic theory including the 
standard model \cite{realistic}, as well as a variant 
dubbed ``racetrack inflation'' \cite{racetrack}.  Following the discussion of
these inflationary models, I will then turn to the issues surrounding the
end of inflation, namely how reheating may be realized \cite{reheating},
the production of cosmic superstring defects \cite{strings}, and the question
of whether our universe itself could be a 3-brane defect created at the end of
inflation \cite{BC}.

\section{Brane-antibrane inflation in flux compactifications}
Intuitively it is quite appealing to imagine that inflation could have
been driven by the potential energy which gives rise to the attractive
force between a brane and antibrane in the early universe \cite{DT,BMNQRZ},
and that reheating results from their mutual annihilation.  However it was
initially not easy to get this simple idea to work in detail.  First, the
potential was not automatically flat enough to get inflation, so it was
necessary to look for means to get around this problem \cite{BMQRZ,JST}.
Even then, there remained a serious shortcoming: various moduli parametrizing
the size and shape of the compact dimensions were considered to be fixed
without specifying the mechanism.  It was not clear whether an actual
stabilization mechanism would preserve whatever degree of flatness one had
managed to achieve for the inflaton potential.

The problem of modulus stabilization received a big boost within the context of
type IIB string theory through reference \cite{GKP} (GKP), where the paradigm of
stabilization by fluxes of NS-NS and R-R gauge fields in 3-cycles of the 
Calabi-Yau manifold appeared (figure 1).  A nice feature of this approach
is that a warped throat with an exponentially large hierarchy could be
generated from a small hierarchies in the ratio of the fluxes.  Such
throats provide a way of solving the hierarchy problem \cite{RS} for an
observer living on a brane in the bottom of the throat.  In this
compactification the dilaton and complex structure (shape) moduli are 
stabilized, but not the overall volume (K\"ahler) modulus.
\centerline{\includegraphics[width=0.6\hsize,angle=0]{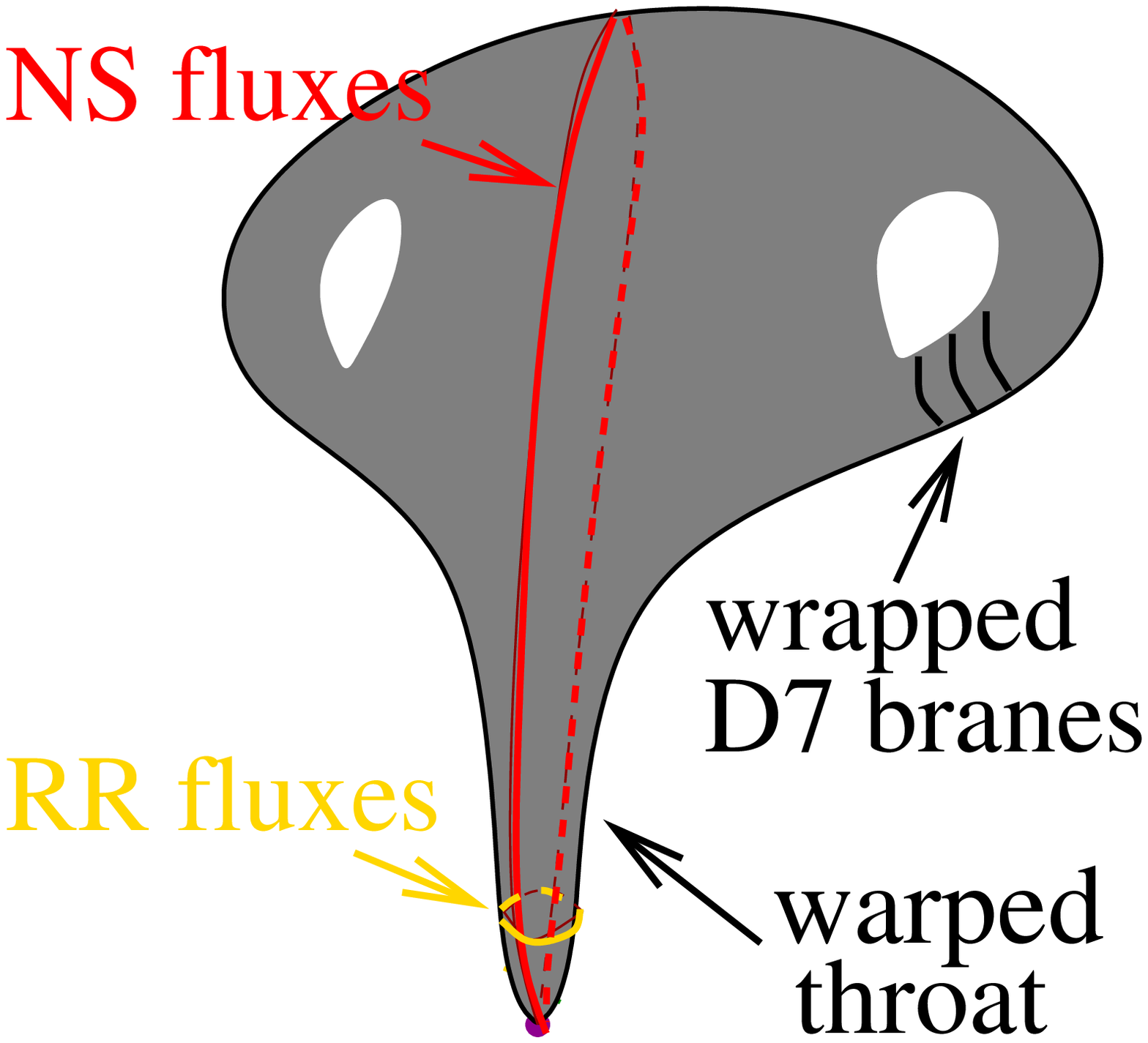}}
\centerline{\small Fig.\ 1: stabilization by fluxes}
\medskip

The GKP idea was put into a cosmological context in reference \cite{KKLT},
where nonperturbative effects like gaugino condensation were invoked to generate
a superpotential for the K\"ahler modulus, $T$,
\beq
\label{W}
	W = W_0 + A e^{-aT}
\eeq
where the constant piece $W_0$ was already arising within GKP.  This
gives rise to a supergravity F-term potential 
\beq
\label{VF}
V_F = e^K\left( K^{a\bar b}D_a W\overline{D_b W} - 3|W|^2\right)
\eeq
with a supersymmetric AdS (negative
energy) minimum, where $T$ is stabilized.  For cosmological purposes one
would like to raise this minimum to a positive value to obtain deSitter space.
KKLT realized that this could be achieved by putting an anti-D3 brane, which
breaks SUSY, into the throat.  The antibrane gives an additional contribution
to the potential,
\beq
\label{VD}
	V_{\bar{D3}} = {k\over T^2}
\eeq
which can raise the minimum of the potential to a positive value.  The 
resulting minimum was only metastable, with the decompactified theory at
$T\to \infty$ being the true minimum, but it was shown that the lifetime
of the deSitter vacuum could easily be cosmologically long.

The step from eternal deSitter space to inflation was made in \cite{KKLMMT}
(KKLMMT) by adding to this picture a mobile D3 brane which would fall down the
throat due to its attraction to the antibrane.  (Note that there should actually
be more than one antibrane in the throat since we want inflation to end with
Minkowski space, not AdS, which would result from the annihilation of the
antibrane if there was only one of them.)  This was the first example of
brane-antibrane inflation where the problem of modulus stabilization was 
addressed in a rigorous way.  The idea would have worked
perfectly, in that the effect of warping in the throat was to make the 
potential of the inflaton (the brane-antibrane separation) naturally small
and thus flat:
\beq
\label{Vinf}
	V_{inf} = {k'\over T^2 |\psi-\psi_0|^4}
\eeq
Here $\psi$ and $\psi_0$ are the positions of the brane and antibrane
respectively, and the coefficient $k'$ is suppressed by exponentially small
warp factors.  However, this nice picture was spoiled by a crucial 
technicality.  Namely, the brane can only be consistently introduced if the
volume modulus appearing in the K\"ahler function $K$ and in the $1/T^2$
factors in (\ref{VD},\ref{Vinf}) becomes $T - f(|\psi|^2)$
instead of just $T$.  The function $f$ can be expanded as $|\psi|^2 + O(\psi^3)$
near $\psi=0$.  Thus the inflaton gets a large additional contribution to its
mass$^2$ which is of order the total potential, $V/M_p^2 \sim H^2$, ruining the
slow-roll condition.  To counteract this, it was necessary to imagine some
source of $\psi$-dependence in the superpotential, which would be fine-tuned
to cancel the unwanted contributions \cite{KKLMMT, BHK, FT}.

\section{Realistic brane-antibrane inflation}

One might also wonder where the standard model (SM) is living in this picture.
In \cite{realistic} we considered a construction in which the SM brane is at a
singular fixed point of a 4-torus in the bottom of the throat, which has been
orbifolded under a $Z_N$ symmetry (figure 2).  These singularities have the advantage of
admitting chiral gauge theories and are thus good candidates for placing the
SM. Each singularity (of which there are many) bring with it a new a modulus, a
blowing-up mode described by a chiral field $B_i$.   We were interested to see
if these modes could have a significant effect on inflation.  To investigate 
this, we simplified the problem by keeping only one of these modes.  

\centerline{\includegraphics[width=0.55\hsize,angle=0]{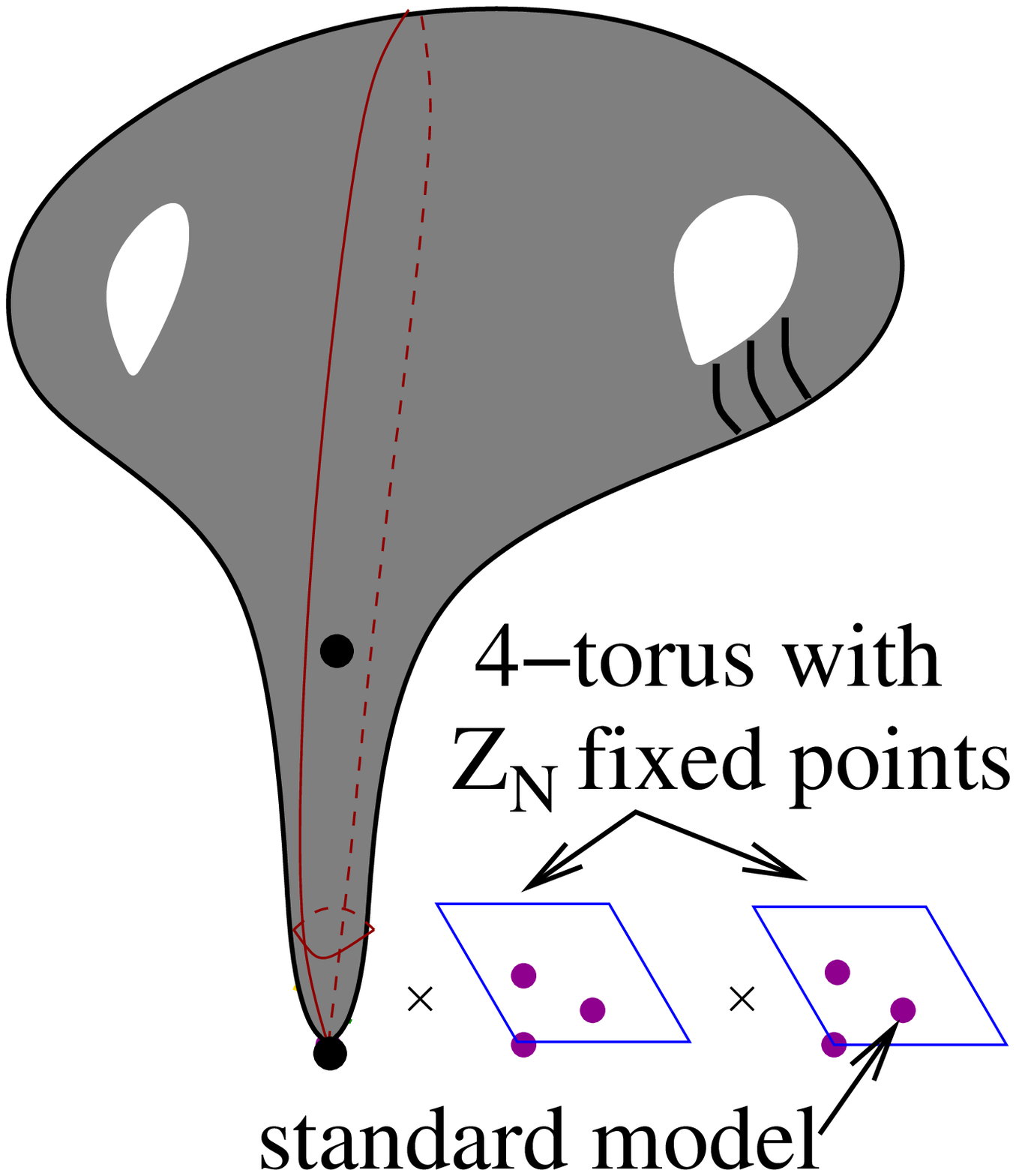}}
\centerline{\small Fig.\ 2: the KKLMMT construction with the Standard Model}
\medskip

The resulting potential is complicated, and depends on three complex fields,
the brane position $\psi + i\phi$, the K\"ahler modulus $T=\sigma+i\tau$
and the blowing-up mode $B+i\beta$:
\beqa
\!\!\!\!\!\!\!\!\!  {V} &=&  {e^{2 B^2} \over r^3}\left(
    2\,{\frac {AC \big( ra +4\,{B}^{2}-2 \,c B \big) \cos \left( \tau -
    \beta \right) }{{e^{a \sigma/2}}{e^{c B}}}}+2\,{\frac {A W_0 \left( r a+4
    \,{B}^{2} \right) \cos  \tau  }{{e^{a\sigma/2}}}} \right. \nonumber\\
    &+&4\,{
    \frac {B C W_0 \left( 2\,B-c \right) \cos \beta }{{e^{c B}}}
    } \left. +{\frac {{A}^{2} \left( 12\,{B}^{2}+6\,r a+r\sigma\,{a}^{2}
     \right) }{3{e^{a\sigma}}}}+{\frac {{C}^{2} \left( 2\,B-c \right) ^{2}}
    { \left( {e^{c B}} \right) ^{2}}}+4\,{B}^{2}{W_0}^{2} \right) \nonumber\\
&+& {2{B^2\over f_0^2}} + {{k\over r^2}} + 
{{k'\over r^2|\psi-\psi_0|^4}}
\eeqa
where $r = \sigma - \psi^2$.  The kinetic energy is also complicated. 
However one finds that most of the fields are heavy and can be integrated out,
with the exception of $\sigma$ and $\psi$.  In principle the $\sigma$ field
could also be integrated out, but since its nontrivial minimum is a delicate
feature, we keep track of its dynamics.   In the right regime of parameters, 
the potential looks like figure 3, with a trough in the $\psi$ direction at
the value of $\sigma$ where it is stabilized.

\centerline{
\includegraphics[width=0.5\hsize,angle=0]{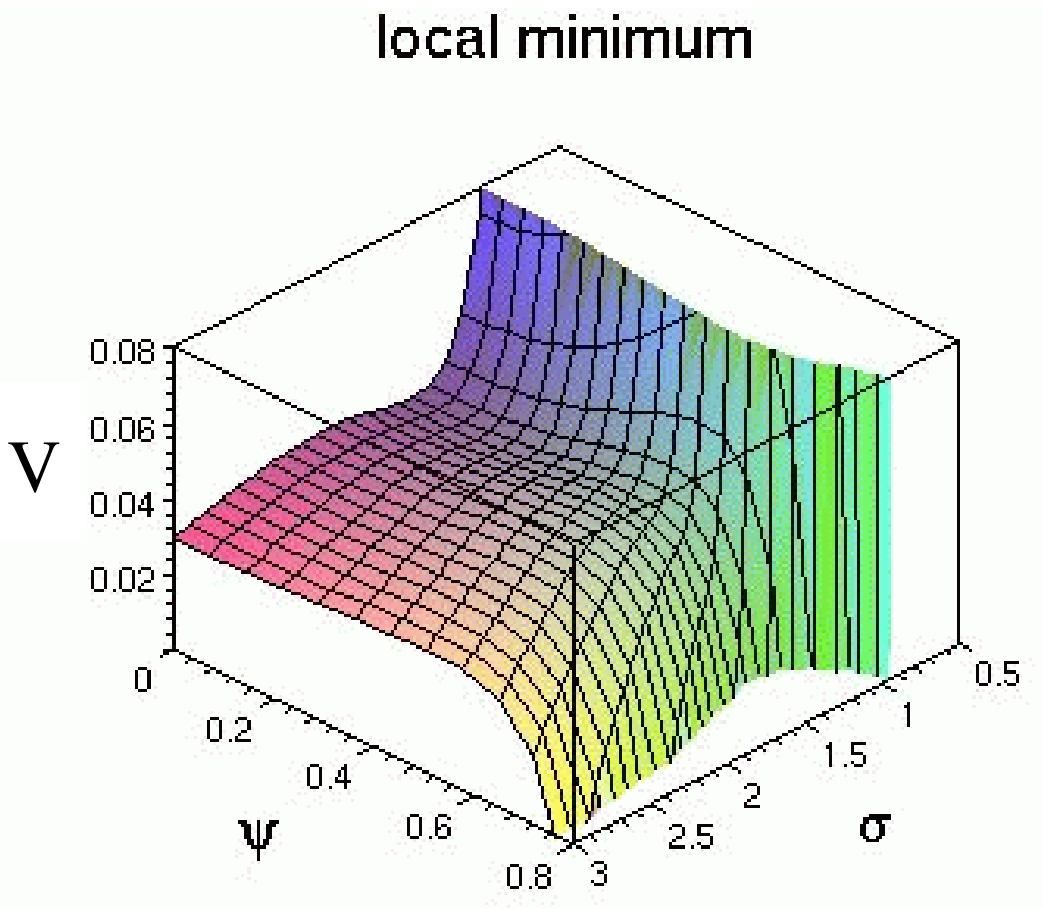}
\hfil\includegraphics[width=0.5\hsize,angle=0]{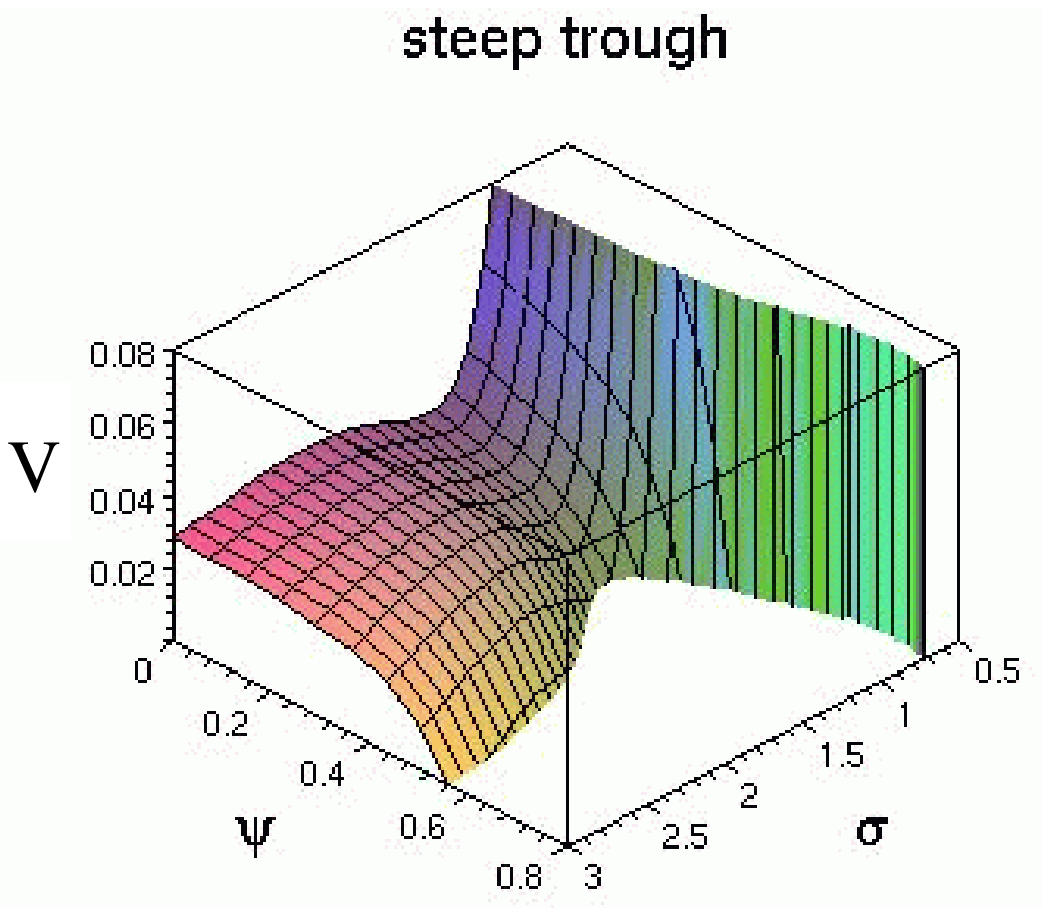}
}
{\small Fig.\ 3: Inflaton potential not yet tuned to be sufficiently
flat.  There is a local minimum (left) leading to eternal deSitter space, 
or else it rolls too quickly to give enough inflation (right).}
\medskip

We found that it is possible to tune parameters so that the trough is
sufficiently flat, without introducing any new source of $\psi$ dependence such
as in the superpotential.  Hence this is in way a simpler realization than was
suggested in KKLMMT.  Detailed study of the inflaton dynamics shows that
the degree of fine-tuning needed is rather severe however.  Figure 4 shows
that the value of $k$ needed to get a given number of e-foldings of inflation,
for some fixed values of the other parameters, must be tuned to a part in 1000
to get 60 e-foldings.

\centerline{$\!\!\!\!\!\!\!\!\!$\includegraphics[width=0.56\hsize,angle=0]{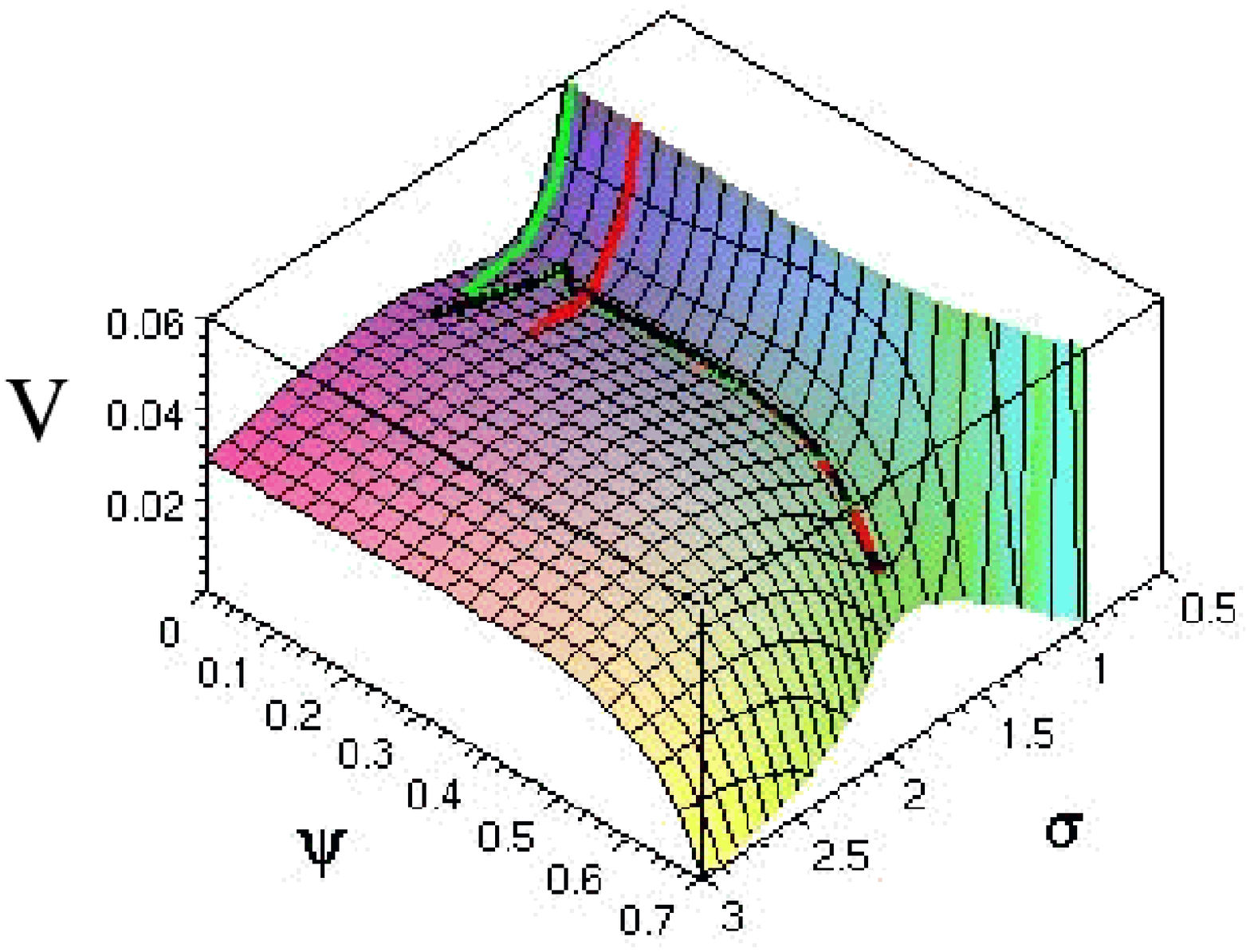}
\hfil
\raisebox{0.275cm}{\includegraphics[width=0.5\hsize,angle=0]{k.eps}}
}
{\small Fig.\ 4: To obtain a flat trough (left), strong tuning of parameters is
required (right).} 
\medskip

Nevertheless, other models of inflation also require tuning to get a flat 
potential, so it is worthwhile to reserve judgment and consider whether the
theory has any distinctive predictions.  One prediction is that the scalar
spectral index may have significant deviations from $n=1$ and some level of
negative running, $dn/d\ln k < 0$, as illustrated in figure 5, for a model
where the COBE fluctuations are produced between 30 and 40 e-foldings after the
beginning of inflation.   

\centerline{
\raisebox{3.05cm}{\parbox{0.5\textwidth}
{\includegraphics[width=\hsize,angle=0]{index-closeup.eps}]}}
\hfil\includegraphics[width=0.482\hsize,angle=0]{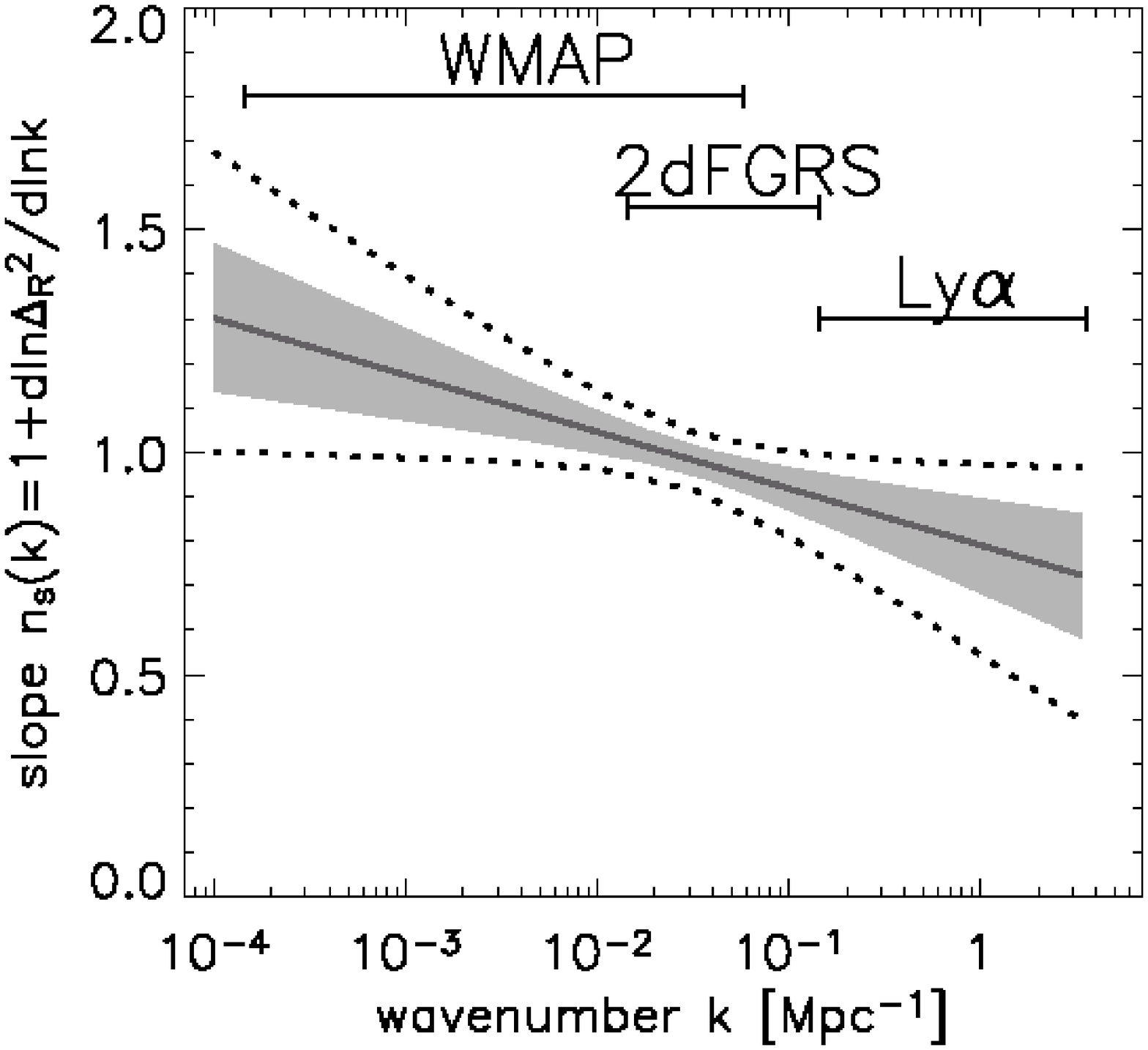}}
{\small Fig.\ 5: Left: deviation of scalar spectral index from 1 as a function of $N\sim \ln k$, over the
range visible in the CMB for a particular choice of model parameters.  Right: WMAP, 2DF galaxy redshift survey and
Lyman alpha combined constraints.}

\vfil

Unfortunately such predictions are only indicative and not robust, since
one can always tune the potential to be even more flat, to further suppress
the deviations from $n=1$. However the fact that more tuning would be required
encourages one to believe that the most natural situation is one in which
inflation lasted for only a short time (the minimal canonical 60 e-foldings),
which leads to the largest deviations from $n=1$ which are not yet ruled out.
Interestingly the constraints are becoming stronger with the combination of
Lyman alpha forest data from the Sloan Digital Sky Survey with WMAP data
\cite{SDSS}, claiming a determination of $n=0.98\pm 0.02$.  

It is also interesting that one can find more exotic inflaton trajectories
in which the K\"ahler modulus plays a role, by opening up a gap in the
trough through which $\sigma$ can escape, as shown in figure 6.  (To be
realistic, these would require additional metastable minima at larger
values of $\sigma$ to prevent decompactification of the extra dimensions.)  
Such complications could give rise to isocurvature fluctuations, and to
bumpy features in the scalar spectrum as shown in figure 7.  Although there
is not yet evidence of such features, we can hope they will appear in future
data, giving a valuable fingerprint that will help to identify which
inflationary model is correct.
\vfil

\centerline{
\includegraphics[width=0.5\hsize,angle=0]{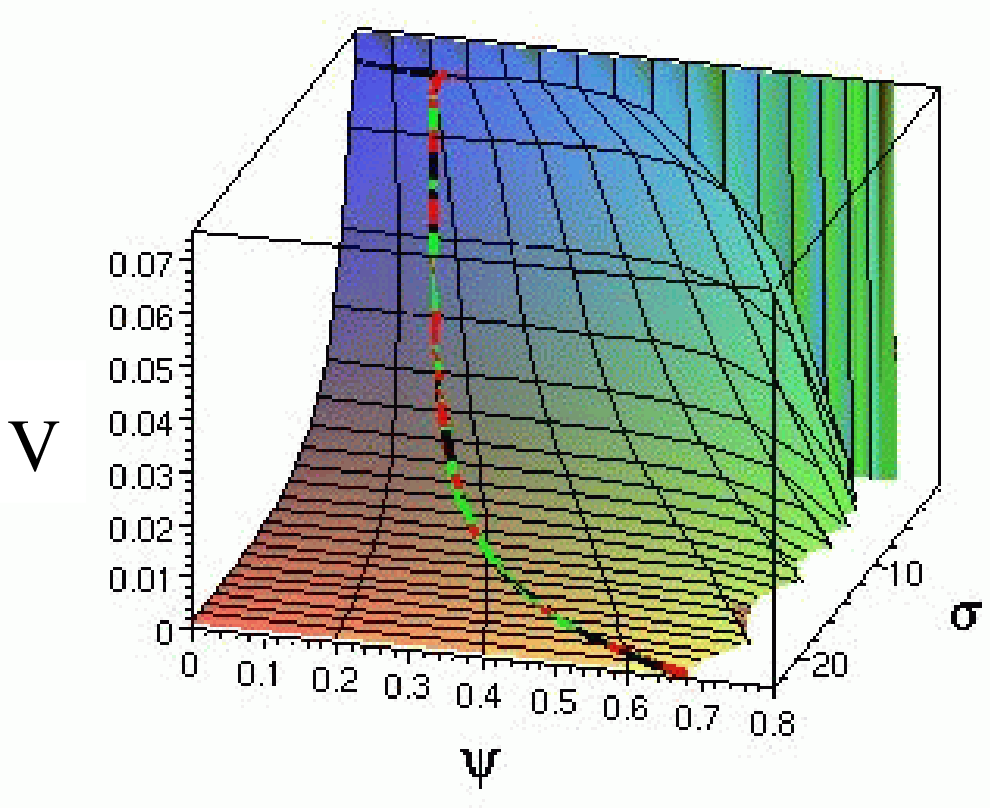}\hfil
\includegraphics[width=0.5\hsize,angle=0]{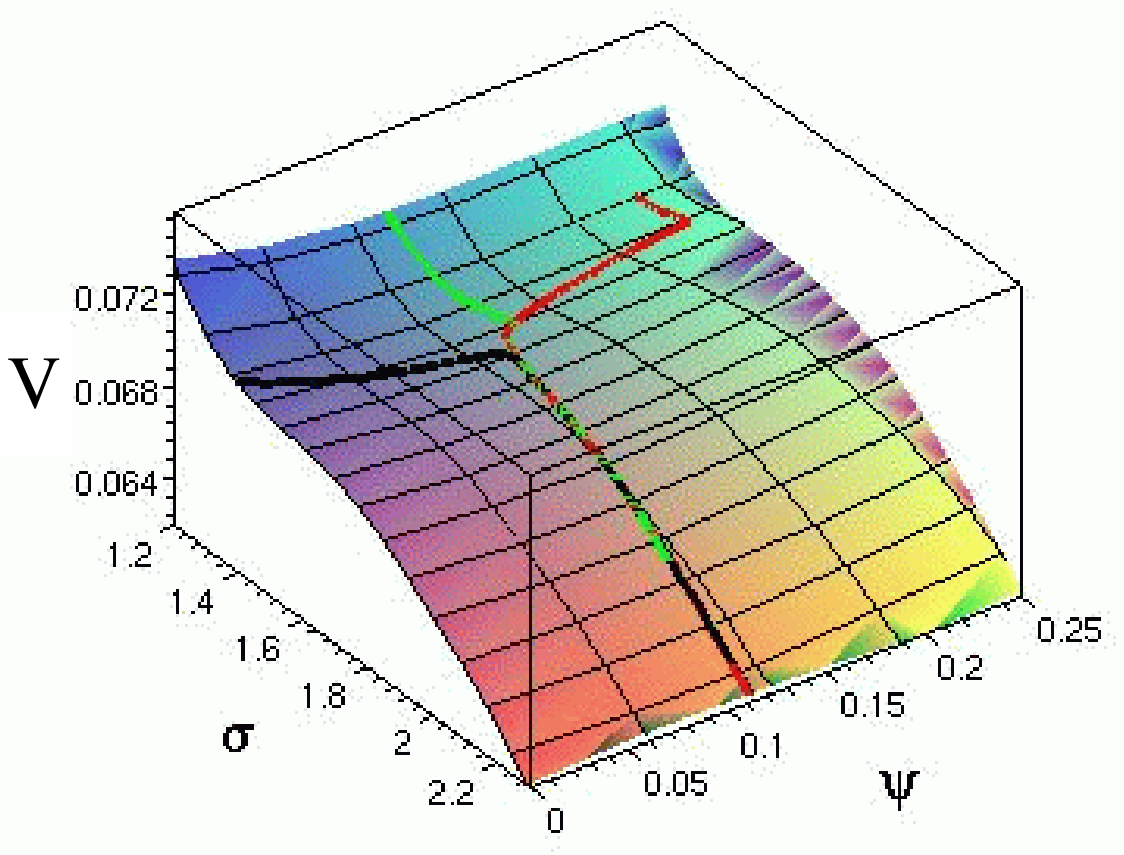}}
{\small Fig.\ 6: Examples of inflation with combined motion of brane and 
K\"ahler moduli.}

\newpage

\centerline{$\!\!\!\!\!\!\!\!\!\!\!$
\includegraphics[width=0.5\hsize,angle=0]{dndlnk2.eps}\hfil
\includegraphics[width=0.5\hsize,angle=0]{dndlnk4.eps}}
\smallskip
{\small Fig.\ 7: Features in the power spectrum arising from the twisted inflaton
trajectories of fig.\ 6, right.} 

\section{Racetrack inflation}

In \cite{racetrack} we considered a variant of the KKLMMT model which is
simpler, in that no mobile D3 brane is required.  Instead we enlarge the
gauge group in which the gauginos are condensing to be a product,
SU(N)$\times$SU(M), which generalizes the superpotential to
\beq
 W = {W_0} + Ae^{-aT} + B e^{-bT}
\eeq
where $a=2\pi/N$ and $b=2\pi/M$.  
Remarkably, this sufficient to obtain inflation from the K\"ahler field
itself, which marks perhaps the first time where string moduli have been
successfully used for inflation in a rigorous setting.  The potential
is simply the F-term (\ref{VF}) and the antibrane term (\ref{VD}).  The
former by itself would be the conventional racetrack potential which has
been previously studied for modulus stabilization.

Writing $T = X + i Y$, the potential can be viewed as in figure 8.  Since
it has two degenerate minima, it provides an example of topological inflation:
there will be separated domains of the universe in different minima, and
between them inflating domain walls which guarantee the existence of regions
in the universe where the initial conditions are correct for inflation to take
place.

\centerline{\includegraphics[width=0.7\hsize,angle=0]{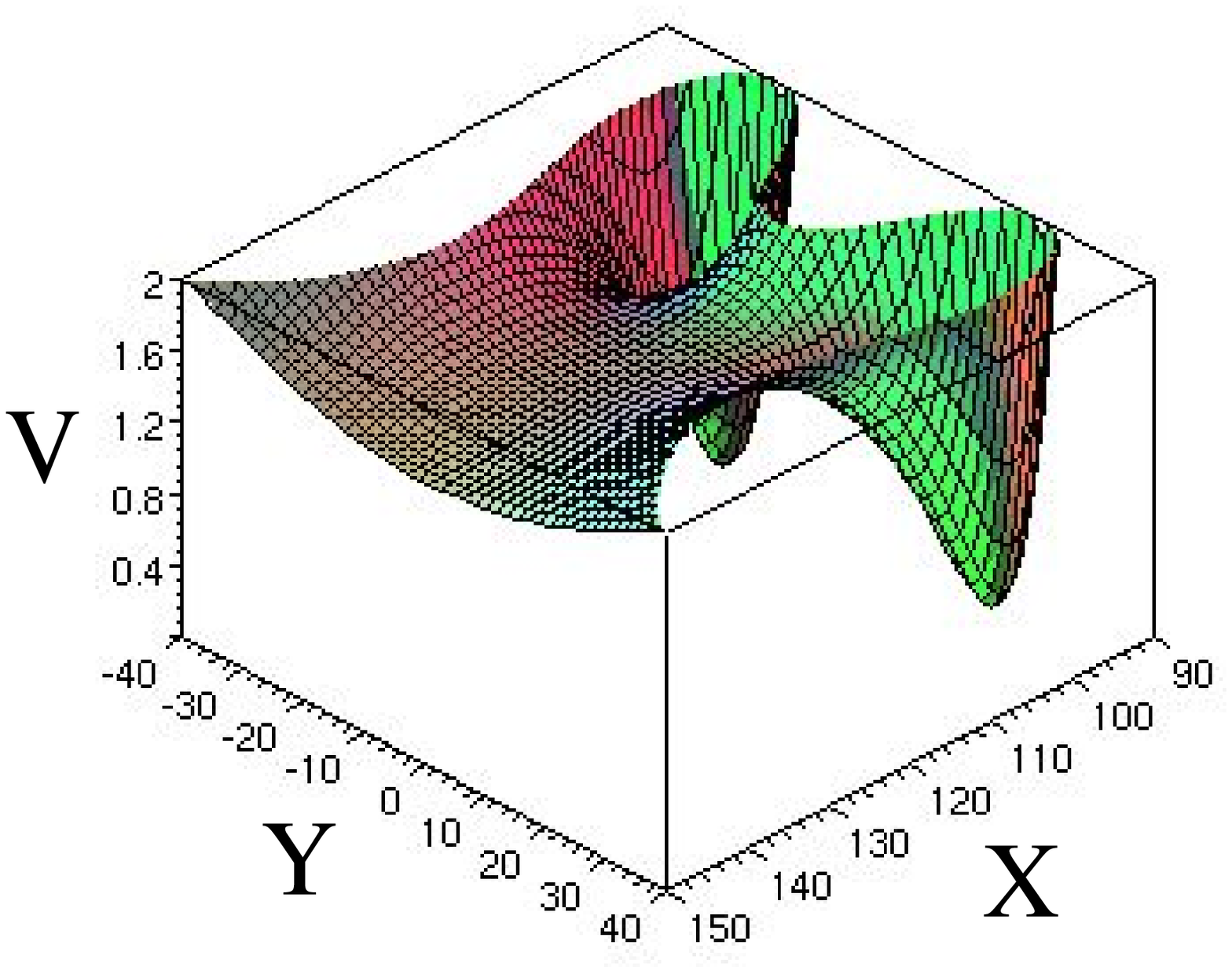}}
\centerline{\small Fig.\ 8: The racetrack inflation potential}
\newpage

Unfortunately this model is just as fine tuned as the previous one.
For sample values of the parameters given by 
$A=\frac{20}{1000}, \ B=-\, \frac{34}{1000}, \
a=\frac{2\pi}{100}, \  b=\frac{2\pi}{90}$, the term $W_0$ must lie in the
narrow range $ -\frac{1}{20389} \le W_0 \le -\frac{1}{20400}$, constituting
a 1 part in 2000 tuning.  On the positive side, the predictions of this model
are more robust than in the previous one.  We find that regardless of how
long inflation lasts, the spectral index 60 e-foldings before the end of
inflation is always near $n=0.95$, as shown in figure 9.  It is interesting that
such values may already be starting to feel pressure from the observational
constraints.

\bigskip\bigskip\bigskip
\centerline{\includegraphics[width=0.5\hsize,angle=0]
{pert.eps}}
\centerline{\small Fig.\ 9: spectral index of the racetrack model}

\section{The reheating problem in brane-antibrane inflation}
It is not as obvious how reheating will work in brane-antibrane inflation
as in conventional inflation models.  The brane-antibrane case is almost
like hybrid inflation, but there is an important difference, illustrated 
in figure 10.   In both cases, the end of inflation is signaled by a field
orthogonal to the inflaton becoming tachyonic.  In the hybrid
case this second field rolls to a stable minimum, oscillates around it,
and causes conventional reheating (or preheating).  
In the string-theory case, the tachyonic
field is described by Sen's effective action (for a review see 
\cite{sen}),
\beq
\label{sen}
S = -\int d^{\,p+1}x\, e^{-|T|^2/a^2}\left(\sqrt{1 - |\partial T|^2} \right)
\eeq
Here $a$ is the string length scale, and 
$T$ now stands for the mode of the open string between a brane and
antibrane which becomes tachyonic when their distance becomes smaller than
some critical value.  The potential is minimized only as $T\to\infty$, with
no oscillations that can give reheating in the usual way.  In fact the whole
action vanishes in this limit.  At zero string coupling, the coherent
tachyon condensate would persist forever, with the equation of state of
cold dark matter, and the universe would never reheat.  At nonzero string 
coupling, the tachyon fluid quickly decays into closed strings.  There is 
a danger that these closed string states will naturally decay into the ground
state of the closed string, namely gravitons.  If this is the whole story, then
there is still no reheating, since the gravitons are not visible radiation.

In \cite{reheating} we have pointed out a very simple solution to this problem,
namely that inflation should take place in a separate throat from that where
the SM is living, as depicted in figure 11(a).  This is already a very natural
assumption to make from the standpoint of the hierarchy problem, since large
warping in the SM model throat is needed to explain the smallness of the Higgs
mass, whereas we need a larger energy scale hence smaller warping in the
inflation throat to get the right level of density fluctuations.

To see why two throats help solve the reheating problem, notice that
the geometry of the throats is like that of the Randall-Sundrum model \cite{RS}:
\beq	ds^2 = a^2(y)\, dx^2 + dy^2 + y^2 d\Omega_5^2 \eeq
where $y$ represents distance along the throat and $\Omega_5$ are the orthogonal
directions of the 6D Calabi-Yau space.  The warp factor is an exponential,
$a(y) = e^{-ky}$.  The idea is that the closed string states which result from
the decay of the tachyon condensate have the right quantum numbers to decay
into Kaluza-Klein graviton excitations.  Consider then the wave functions
of these KK states along the $y$ direction.  Schematically they have the form
shown in figure 12: they are exponentially enhanced inside the throats, relative
to outside.  Because of this exponential enhancement, the KK gravitons decay
with the largest branching fraction into particles on branes in the most
strongly warped throats.  The KK gravitons couple to SM model like ordinary 
gravity, but with a suppression of the TeV scale rather than the Planck scale.
Thus efficient reheating of the SM brane is guaranteed so long as it lives in
the deepest throat.

\centerline{\includegraphics[width=0.5\hsize,angle=0]{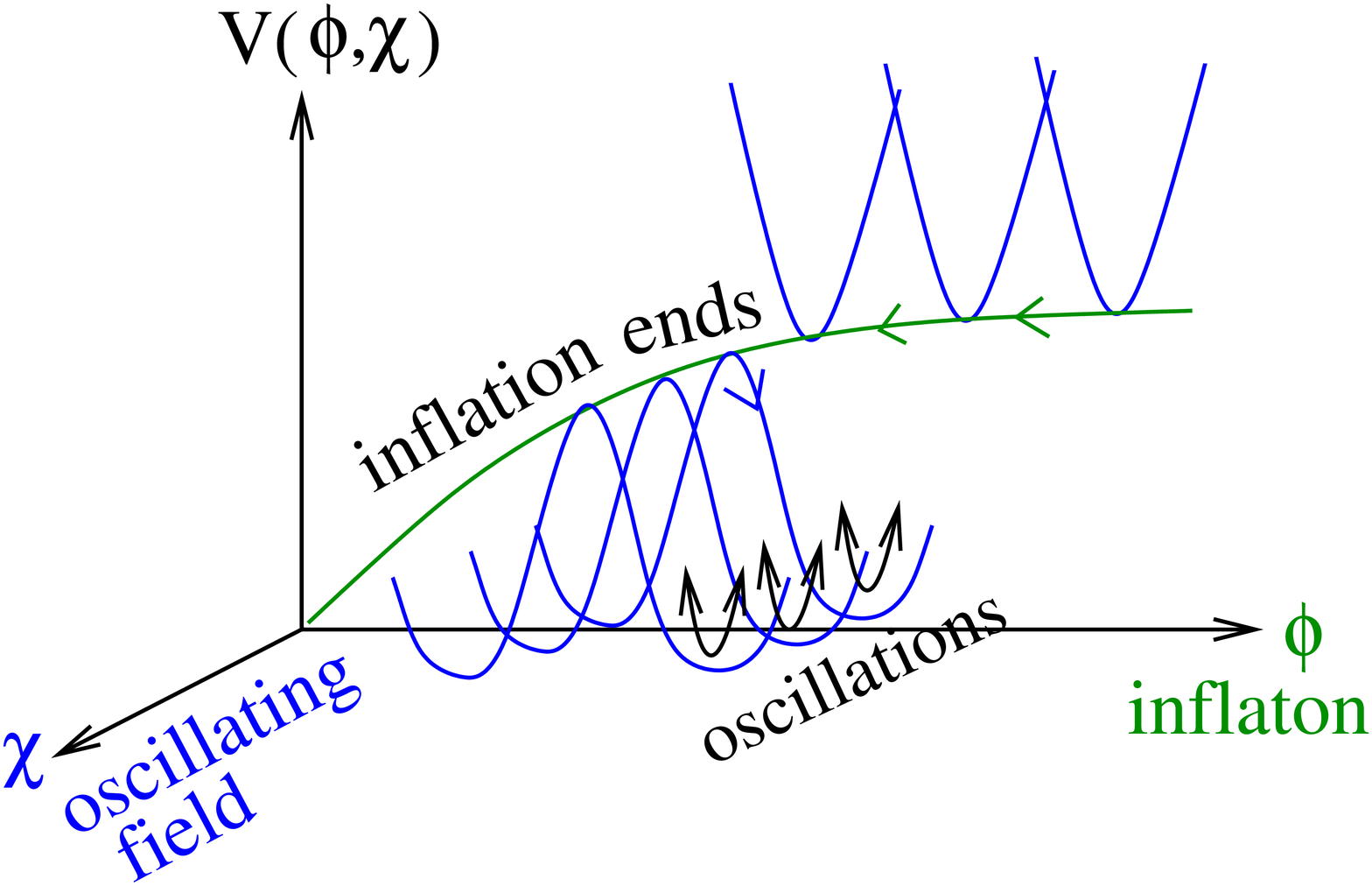}
\includegraphics[width=0.5\hsize,angle=0]{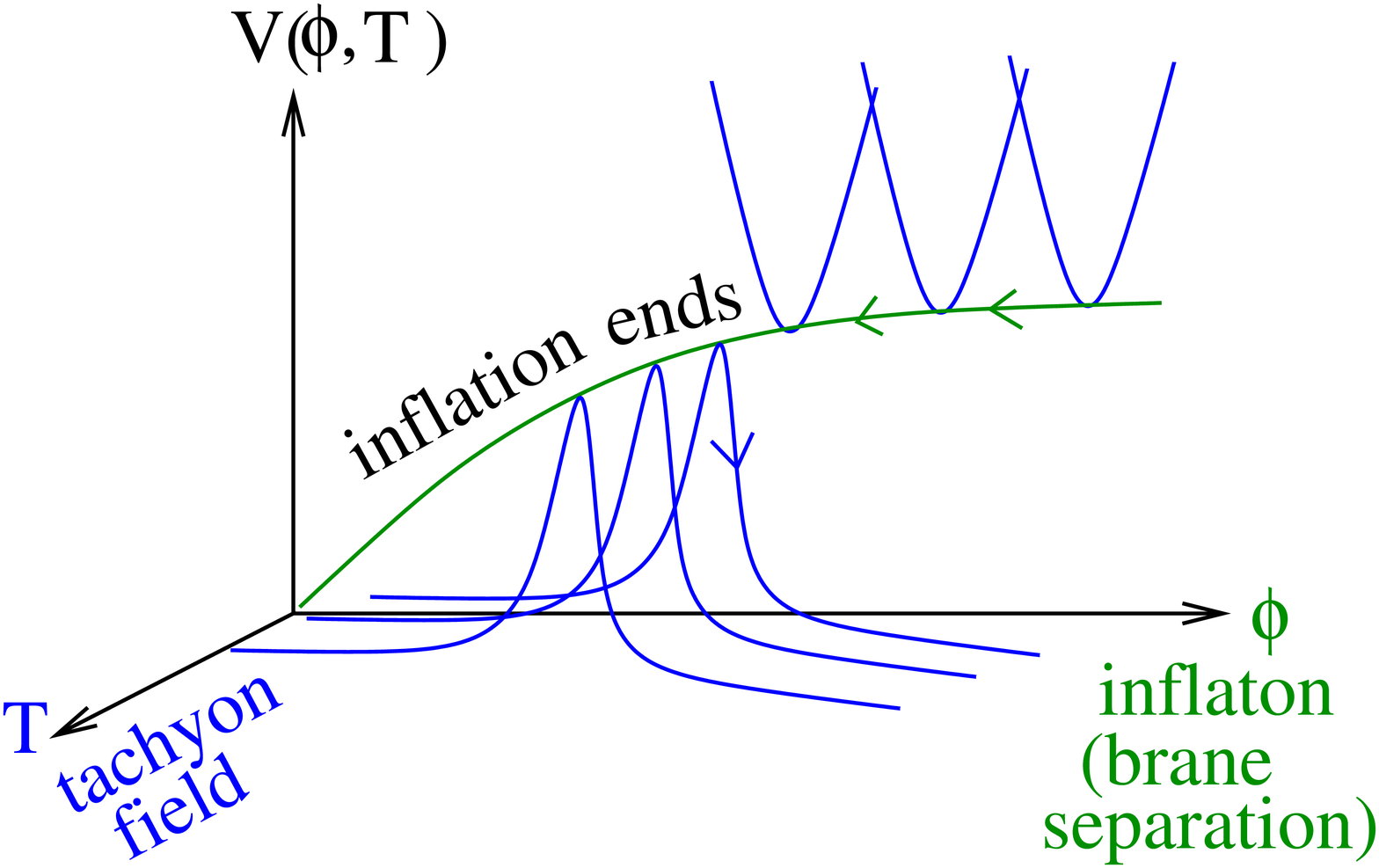}}
\centerline{\small Fig.\ 10: potentials for hybrid inflation (left) and brane-antibrane
inflation (right).}

\centerline{\includegraphics[width=0.5\hsize,angle=0]{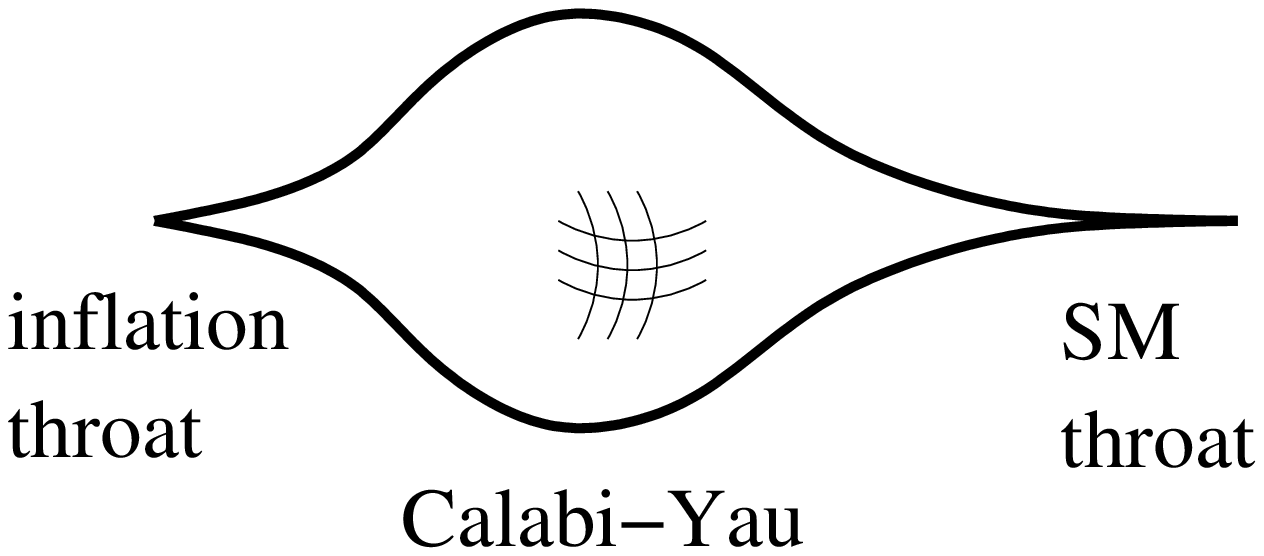}
\includegraphics[width=0.5\hsize,angle=0]{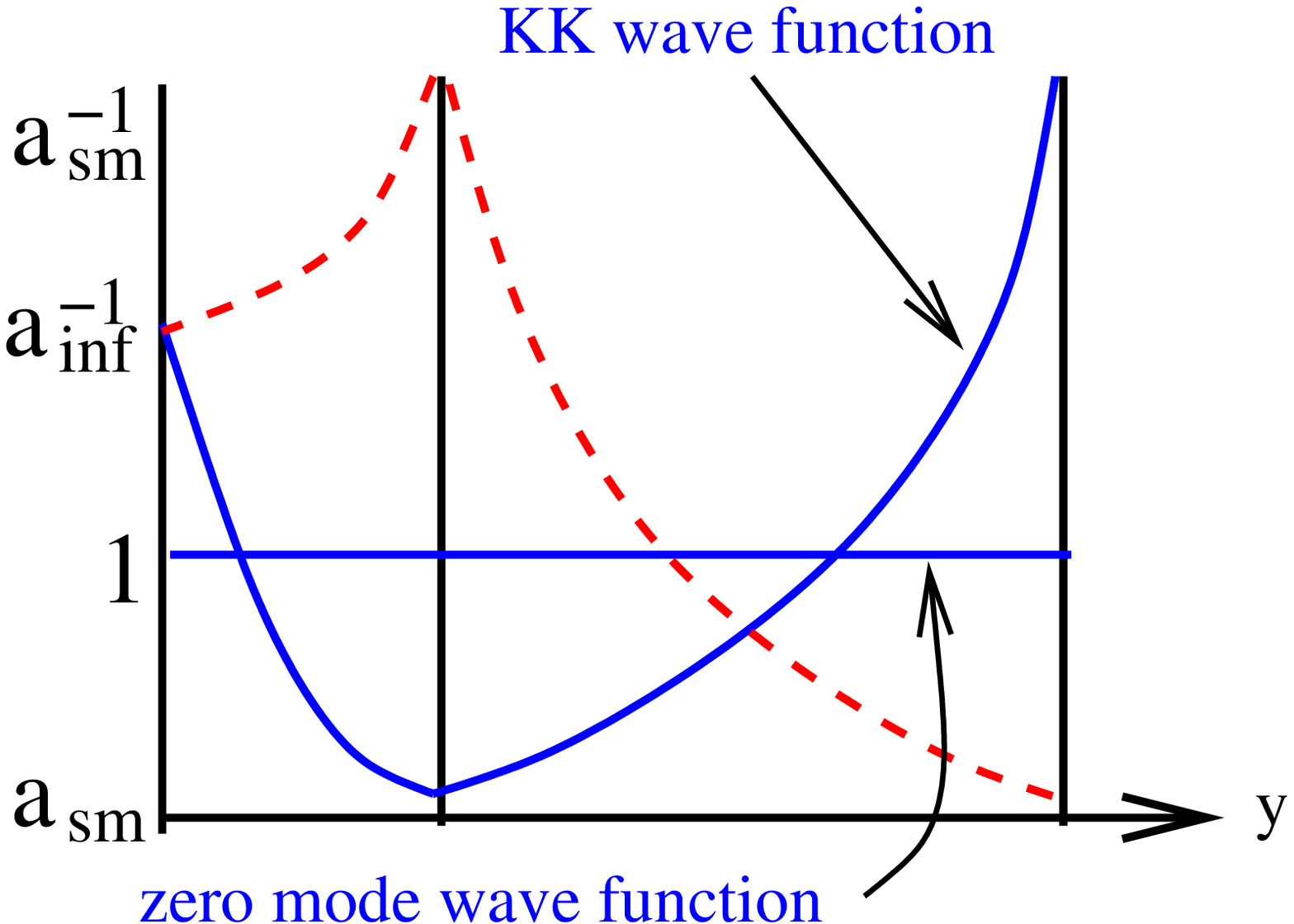}}
{\small Fig.\ 11: (a) the two-throat scenario; (b) schematic view of KK 
graviton wave function (showing envelope but not oscillations)
in the two-throat model, with a fictitious Planck brane between the 
throats. }
\medskip

There is another reason why this scenario for reheating is attractive. 
Recently there has been intense interest in the possibility that superstrings
survive as cosmological remnants \cite{ST,CMP}.  Although they cannot be the
dominant source of density fluctuations, a small contribution could be
detectable in the CMB.  The gravity waves created by kinks and cusps on such
strings could also be observable in future pulsar timing measurements, and
especially the LIGO gravitational wave interferometer experiment \cite{DV}.  It
is very exciting that cosmic superstrings could provide the strongest signal
for LIGO, if the string tension $\mu$ is sufficiently low.  The warp factor of
the inflationary throat provided by  the flux compactification scenario gives
the dial needed for adjusting $\mu$ to low enough scales.  However the stability
of relic superstrings requires that the SM brane (and other branes) be separated
from the remnant strings within the extra dimensions---otherwise the strings
can reconnect to branes and break up.  The two-throat picture we propose for
reheating is just what is needed to also allow for the survival of string
relics.

\section{Overproduction of cosmic superstrings?}

At the end of brane-antibrane inflation, the fate of the false vacuum energy
in the tension of the annihilating branes is determined by the dynamics of the
tachyon field in (\ref{sen}).  Because it is a complex field, it is possible for
defects of codimension two to form.  These are topological defects, where the
phase of the field winds around the core of the defect.  
On annihilating 3-branes, they would
be cosmic D-strings, as illustrated in figure 12.  It has been rigorously
established that the defects which form from unstable branes or annihilating
branes are actually consistent with being D-branes themselves, of smaller
dimensionality \cite{sen}. 
\vskip-0.15in
\includegraphics[width=0.9\hsize,angle=0]{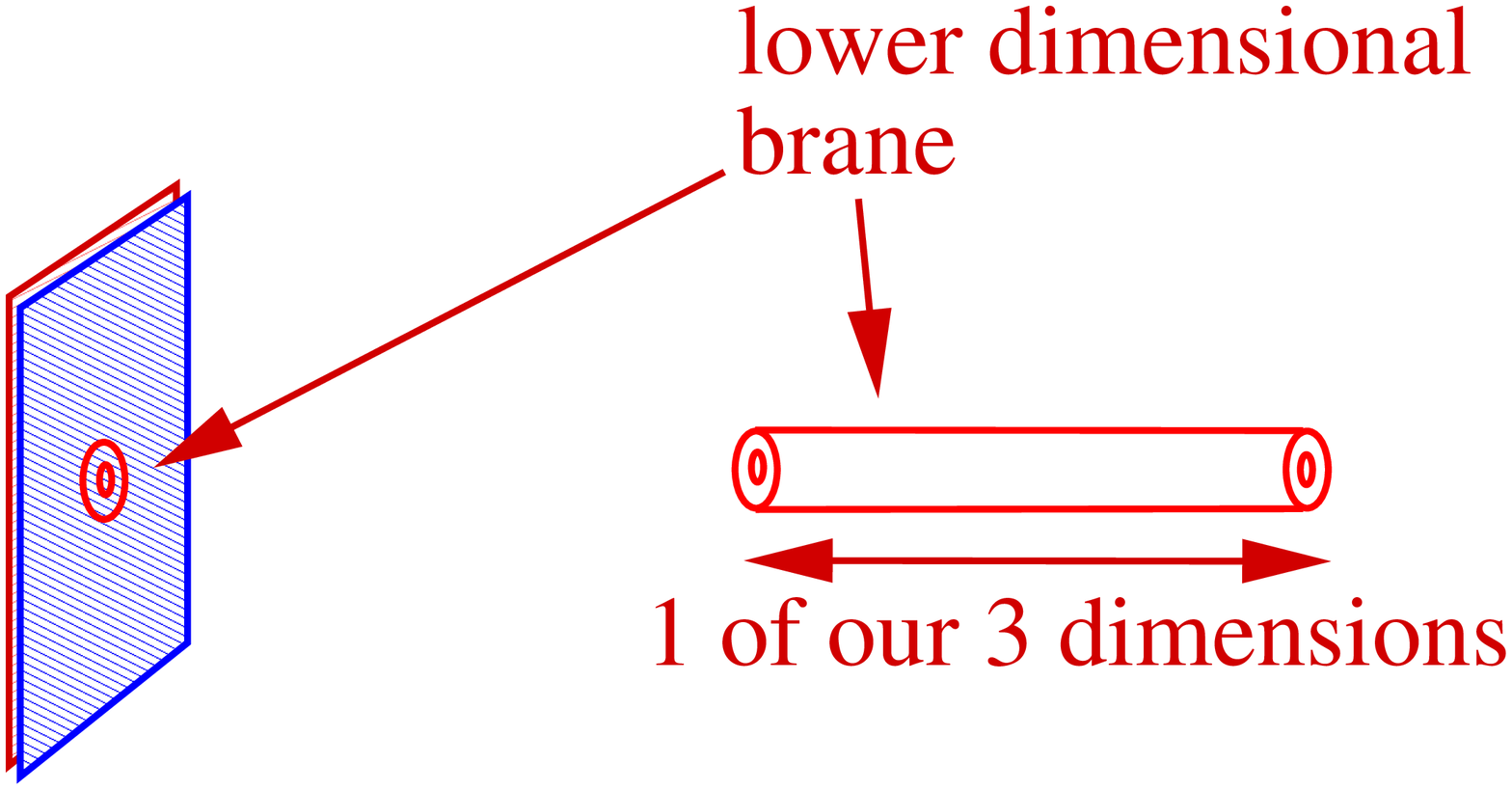}\\
\vskip-0.5in
{\small Fig.\ 12: formation of a cosmic string defect from annihilating 3-branes.}
\medskip 

In the conventional formation of string defects, as in a GUT phase transition,
the initial density of would-be defects is set by a microphysical scale, the inverse
mass of the fields forming the defect.  But in these theories the large initial
gradient energy can be minimized by unwinding of the phase.  This occurs through
the annihilation of nearby defects with opposite winding number.  Such unwinding
can occur between defects that are in causal contact, but in an expanding 
universe, strings in different Hubble volumes cannot interact.  This is the
basis for the Kibble mechanism, guaranteeing at least of order 1 defect
surviving per Hubble volume during the initial stages of defect formation.  
It is worth noting however that this is only a lower bound.  If the defects were
less efficient in annihilating, many more could survive this initial stage of
unwinding.  

The unusual form of the action (\ref{sen}) leads to precisely to such an
expectation for cosmic superstrings, that their initial density will be much
higher than the minimum dictated by Kibble's argument.  This is because of the
exponential suppression factor $e^{-|T|^2/a^2}$ which multiplies the kinetic
term as well as the potential.  Because of this, the restoring force which
would normally attract defects to antidefects quickly becomes negligible
as the tachyon field rolls from its unstable maximum.  As a result, we observe
a high density of initial defects forming from the vacuum quantum fluctuations 
of the $T$ field, as shown in figure 13.

\centerline{\includegraphics[width=0.5\hsize,angle=0]{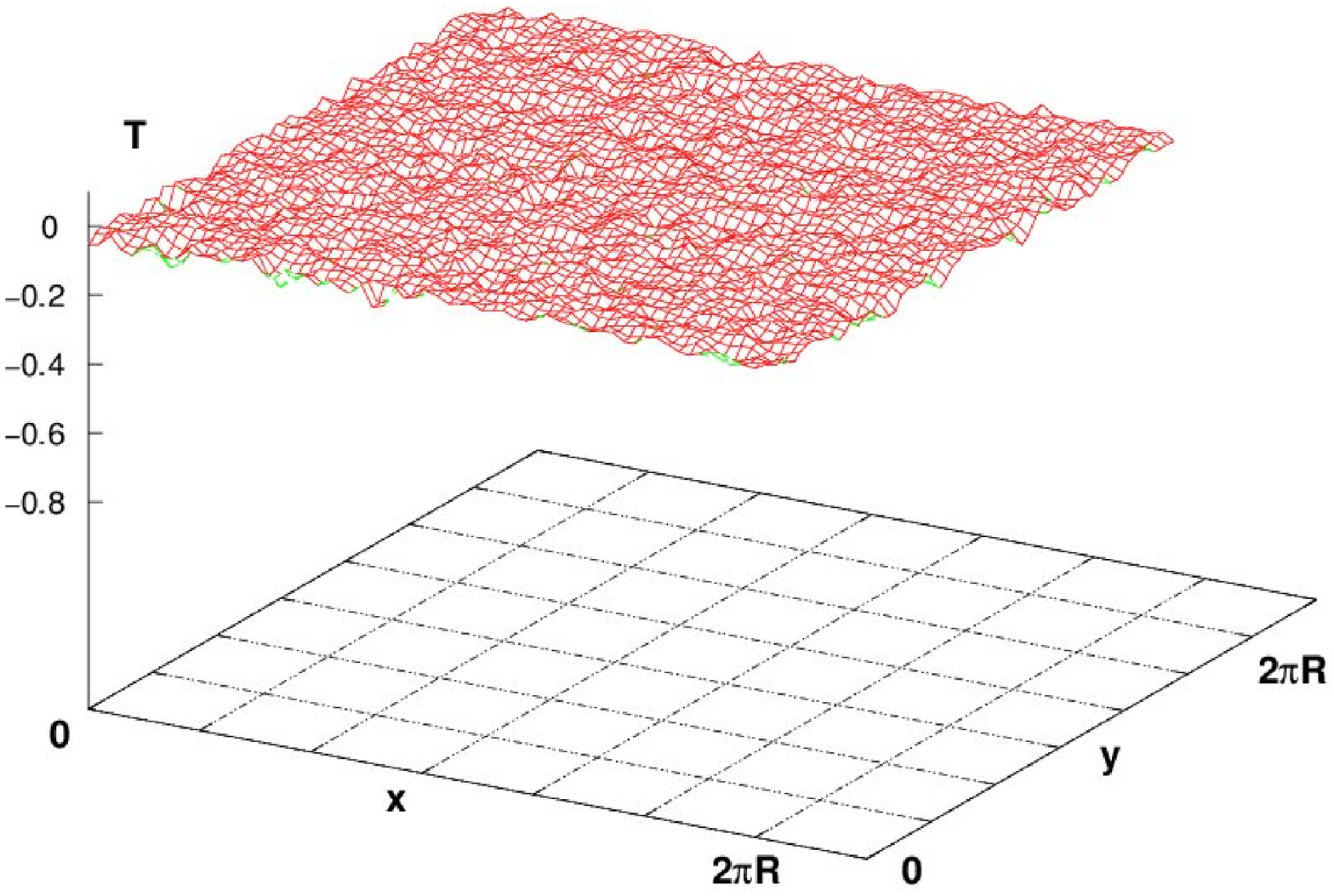}
\includegraphics[width=0.5\hsize,angle=0]{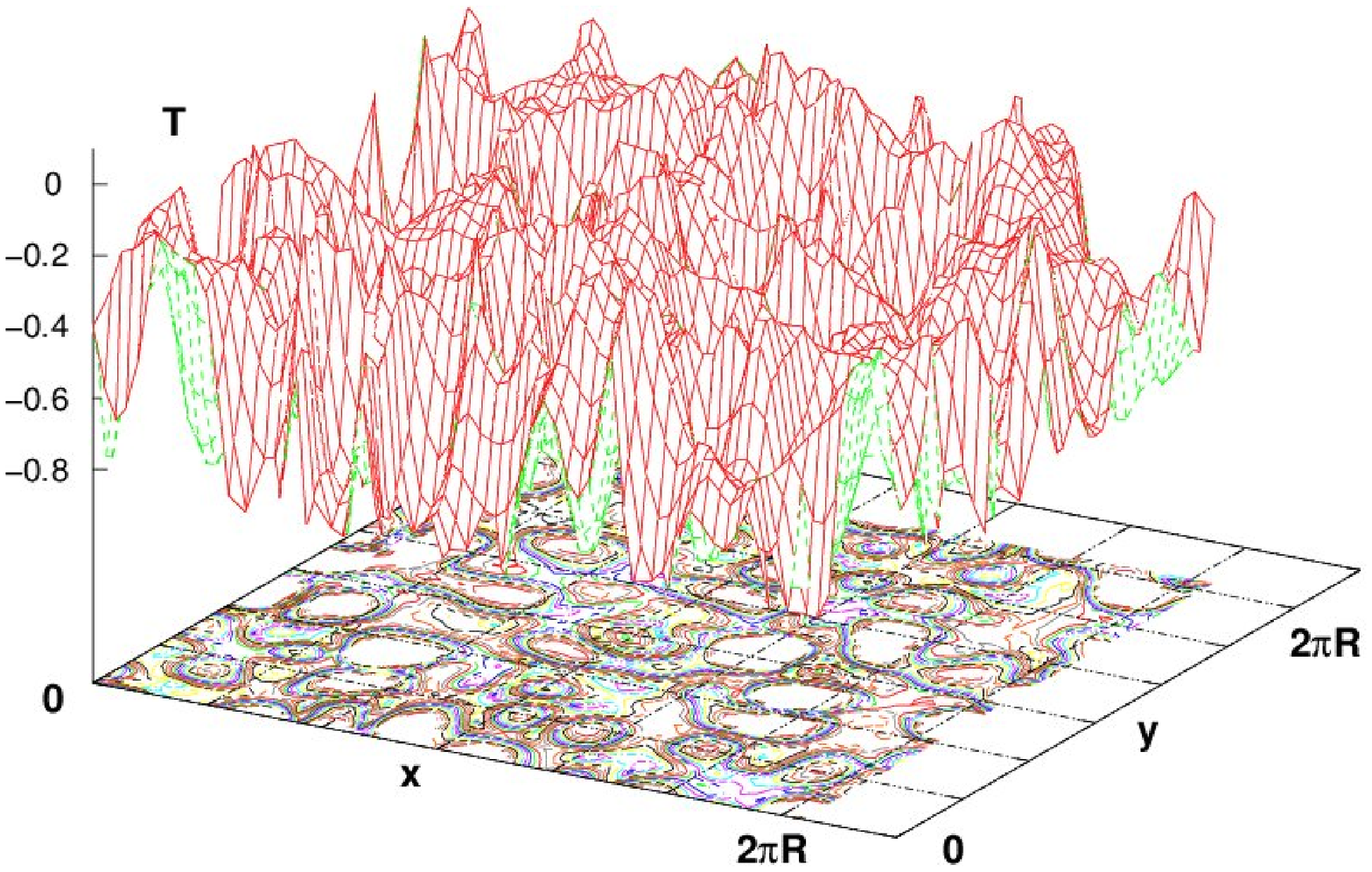}}
{\small Fig.\ 13:  Growth of defects from initial fluctuations for the
open string tachyon field.  Distances are in units of the string length scale.}
\medskip

Even though this initial string density is many orders of magnitude greater than
the Kibble argument lower bound, there appear to be no observational
consequences, thanks to the scaling property of string networks.  In the subsequent
evolution, long strings start to self-intersect, chopping off small loops which
decay away by gravitational radiation.  The steady-state solution is one where
the energy density in strings scales with that of radiation.  We have
investigated the approach to the scaling solution and found that scaling sets
in after a few Hubble times, regardless of how overdense the initial network
was.  This is illustrated in figure 14(a).  

\centerline{\includegraphics[width=0.5\hsize,angle=0]{omega4.eps}
\includegraphics[width=0.5\hsize,angle=0]{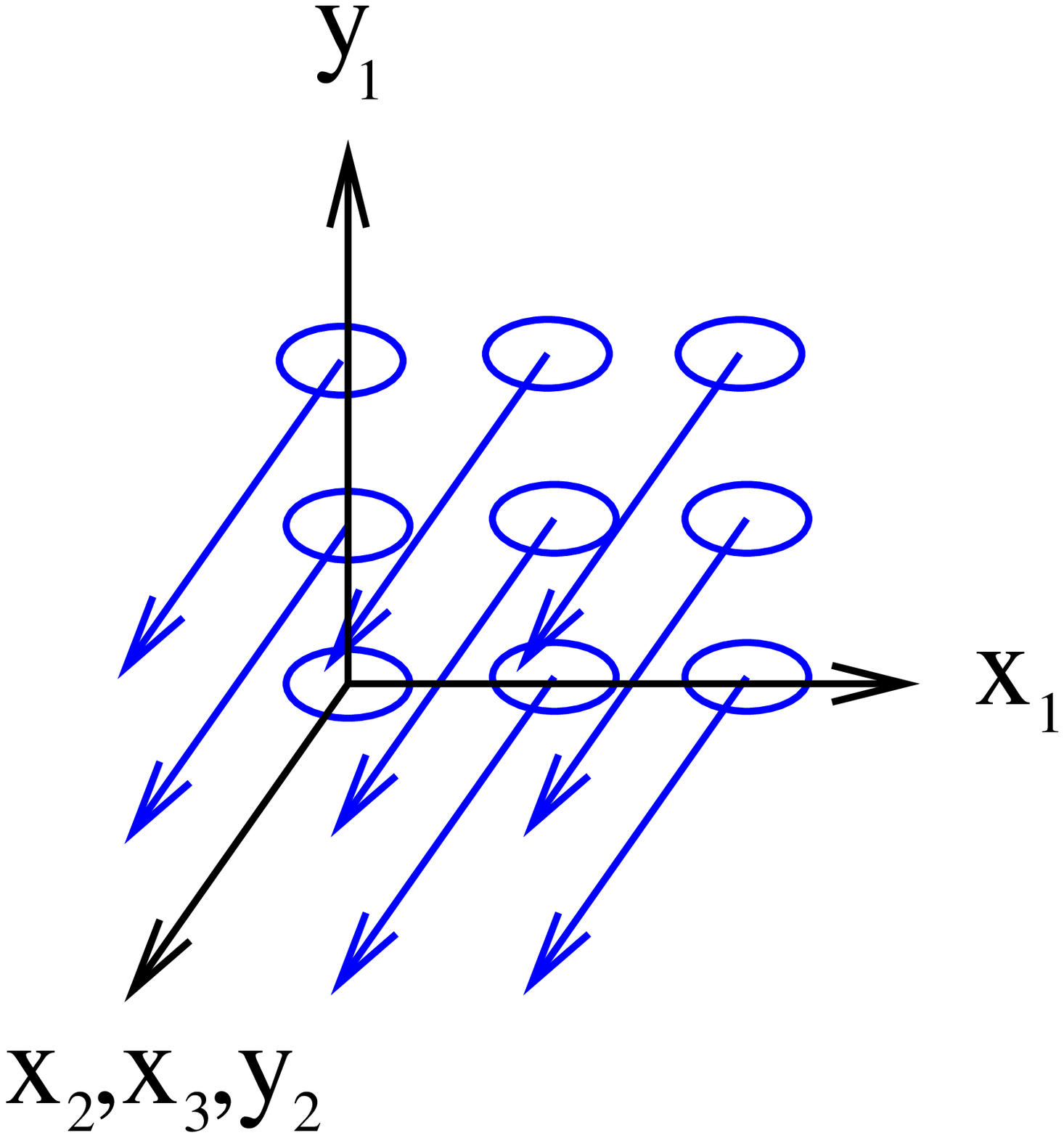}}
{\small Fig.\ 14: (a) Evolution of different components of the energy density
in a string network, starting from equal parts of energy in strings and
radiation.  Not shown is the gravitational radiation component.
(b) Schematic illustration of domain walls forming in large dimensions 
$x_2,x_3$
}\medskip

However, there is one kind of situation where the initial overdensity can make 
an important difference.  This is when the original annihilating branes were
higher than 3-dimensional.  For example, in the annihilation of 5-branes,  the
resulting defects will be 3-branes.  However, the original 5-branes had to wrap
two compact dimensions, so the defects may appear to a 3D observer to be
strings, domain walls, or 3-branes, depending on how they are oriented with
respect to the compact directions.  This is illustrated in fig.\ 14(b) for the
case of domain wall production.   Domain walls are highly constrained relics
since they quickly come to dominate the energy density of the universe.  Unlike
strings, they do not generically shed their energy by self-intersections (which
would pinch  off a closed surface that could decay through gravitational
radiation).  Therefore the defect overproduction problem would at first sight
appear to rule out inflation driven by higher-dimensional branes.  One might
question this conclusion based on the fact that the defects in question are
always close to each other in one of the small compact directions.  Would this
not cause them to annihilate efficiently?  But they can be separated from each
other widely in the large transverse direction, which is growing with the Hubble
expansion.  We therefore consider it an open question, subject to more detailed
simulations of domain wall network evolution in the case where one of the
dimensions tranverse to the walls is compact and small.  

\section{Could our universe be a defect?}

The previous discussion also has implications for another idea for reheating
at the end of brane-antibrane inflation.  It was suggested in \cite{CFM},
and more convincingly argued in \cite{BC}, that our universe could have arisen
just as described above: a 3-brane defect created at the end of 5-brane
annihilation.  In this case the directions transverse to the defect are all
compact ones, so it should be easier for defects to annihilate within this 
compact space.  One can imagine getting around this objection within a
multi-throat picture, where defects and antidefects fell into different throats
and were thus preserved from annihilating each other.  And in the previous
section we overcame another argument against the idea, namely the prejudice that
defects can only form in regions larger than the Hubble radius, whereas the 
compact dimensions are smaller than $H^{-1}$.  The previous analysis shows that
in fact many such defects can initially form within the small dimensions,
provided they are at least a few string lengths in extent.

The main motivation for this picture was that radiation within one of the final
state branes would be very efficiently produced by its coupling to the
time-dependent background tachyon field.  This can be seen by looking at Sen's
action in the presence of U(1) gauge fields originally living on the brane or
antibrane:
\[ S = -\int d^{\,p+1}x\, e^{-|T|^2/a^2}\left(\sqrt{-\det M^{(1)}} 
+ \sqrt{-\det M^{(2)}}
\right)\]
where 
\[ M^{(i)}_{MN} = g_{MN} + F^{(i)}_{MN} + \frac12\left(D_M T\, D_N T^* 
+ {\rm c.c.}\right)\]
\[ D_M = \partial_M -i A_M^{(1)} + i A_M^{(2)} \]

Here the combination $A^{(1)}-A^{(2)}$ gets a VEV, which is part of the defect,
and the orthogonal combination $A^{(1)}+A^{(2)}$ plays the role of a massless
photon.  We find that there is {strong production of $A^{(1)}+A^{(2)}$ 
due to its coupling to the violently changing background $T(t,x)$.
Thus reheating would not be a problem
in such models.  But we are left with the question of why only 3-branes or 
cosmic strings would be produced by 5-brane annihilation, since the domain
walls should appear with equal likelihood, and overclose the universe.  Unless
there is a mechanism to suppress domain walls, we must conclude that it is
unlikely our universe arose as a defect itself, which was created at the end of 
inflation.

\section{Conclusions}

I find it encouraging that we now have  string inflation models which are
starting to come under pressure from the data.  Although not all of the
predictions are hard, some such as those of the racetrack model are.  At last,
a fasifiable prediction coming from string theory!

I would find models of stringy inflation more compelling if they were able
to address the fine tuning problem (for recent progress see \cite{FT,BEMMM}).
On the other hand, the tuning problem tells us that the most natural situation,
which is least fine-tuned, is the one with the minimal amount of inflation
needed to explain what is observed.  This opens the door for us to expect 
nongeneric features in the power spectrum:
departures from flatness, running of the spectral index,
isocurvature fluctuations, bumps and wiggles in the spectrum\dots
Such features would obviously make the interaction between theory and
observation more interesting, and I hope the observers will find them soon.

In this talk I have argued that  two-throat models of brane-antibrane
inflation  naturally solve the problem of reheating, explaining the hierarchy
between the inflation scale and the TeV scale of the standard model, and they
help to preserve possible  cosmic superstring defects.   It is a very exciting
prospect that gravitational waves observed by  LIGO could be first signal of
string theory in the laboratory. Although it may be challenging to distinguish
cosmic superstrings from conventional cosmic strings, in principle this is a good
counterexample to the the naysayers of string theory's testability.

\section{Acknowledgements}
I warmly thank Jiro Soda, Misao Sasaki, Takahiro Tanaka and
the other organizers of JGRG14 for their kind hospitality during the conference.


\begin{thebibliography}{99}


\bibitem{realistic}
C.~P.~Burgess, J.~M.~Cline, H.~Stoica and F.~Quevedo,
``Inflation in realistic D-brane models,''
JHEP {\bf 0409}, 033 (2004)
[arXiv:hep-th/0403119].

\bibitem{racetrack}
J.~J.~Blanco-Pillado {\it et al.},
``Racetrack inflation,''
JHEP {\bf 0411}, 063 (2004)
[arXiv:hep-th/0406230].

\bibitem{reheating}
N.~Barnaby, C.~P.~Burgess and J.~M.~Cline,
``Warped reheating in brane-antibrane inflation,''
arXiv:hep-th/0412040.

\bibitem{strings}
N.~Barnaby, A.~Berndsen, J.~M.~Cline and H.~Stoica,
``Overproduction of cosmic superstrings,''
arXiv:hep-th/0412095.

\bibitem{BC}
N.~Barnaby and J.~M.~Cline,
``Creating the universe from brane-antibrane annihilation,''
Phys.\ Rev.\ D {\bf 70}, 023506 (2004)
[arXiv:hep-th/0403223].

\bibitem{KKLT}
S.~Kachru, R.~Kallosh, A.~Linde and S.~P.~Trivedi,
``De Sitter vacua in string theory,''
Phys.\ Rev.\ D {\bf 68}, 046005 (2003)
[arXiv:hep-th/0301240].


\bibitem{KKLMMT}
S.~Kachru, R.~Kallosh, A.~Linde, J.~Maldacena, L.~McAllister and S.~P.~Trivedi,
``Towards inflation in string theory,''
JCAP {\bf 0310}, 013 (2003)
[arXiv:hep-th/0308055].

\bibitem{DT}
G.~R.~Dvali and S.~H.~H.~Tye,
``Brane inflation,''
Phys.\ Lett.\ B {\bf 450}, 72 (1999)
[arXiv:hep-ph/9812483].

\bibitem{BMNQRZ}
C.~P.~Burgess, M.~Majumdar, D.~Nolte, F.~Quevedo, G.~Rajesh and R.~J.~Zhang,
``The inflationary brane-antibrane universe,''
JHEP {\bf 0107}, 047 (2001)
[arXiv:hep-th/0105204].

\bibitem{BMQRZ}
C.~P.~Burgess, P.~Martineau, F.~Quevedo, G.~Rajesh and R.~J.~Zhang,
``Brane antibrane inflation in orbifold and orientifold models,''
JHEP {\bf 0203}, 052 (2002)
[arXiv:hep-th/0111025].

\bibitem{JST}
N.~Jones, H.~Stoica and S.~H.~H.~Tye,
JHEP {\bf 0207}, 051 (2002)
[arXiv:hep-th/0203163].

\bibitem{GKP}
S.~B.~Giddings, S.~Kachru and J.~Polchinski,
``Hierarchies from fluxes in string compactifications,''
Phys.\ Rev.\ D {\bf 66}, 106006 (2002)
[arXiv:hep-th/0105097].

\bibitem{RS}
L.~Randall and R.~Sundrum,
``A large mass hierarchy from a small extra dimension,''
Phys.\ Rev.\ Lett.\  {\bf 83}, 3370 (1999)
[arXiv:hep-ph/9905221].

\bibitem{BHK}
M.~Berg, M.~Haack and B.~Kors,
``On the moduli dependence of nonperturbative superpotentials in brane
arXiv:hep-th/0409282; 
``Loop corrections to volume moduli and inflation in string theory,''
Phys.\ Rev.\ D {\bf 71}, 026005 (2005)
[arXiv:hep-th/0404087].

\bibitem{FT}
H.~Firouzjahi and S.~H.~H.~Tye,
``Closer towards inflation in string theory,''
Phys.\ Lett.\ B {\bf 584}, 147 (2004)
[arXiv:hep-th/0312020].

\bibitem{SDSS}
U.~Seljak {\it et al.},
 ``Cosmological parameter analysis including SDSS Ly-alpha forest and galaxy
bias: Constraints on the primordial spectrum of fluctuations, neutrino mass,
and dark energy,''
arXiv:astro-ph/0407372.

\bibitem{sen}
A.~Sen,
``Tachyon dynamics in open string theory,''
arXiv:hep-th/0410103.

\bibitem{ST}
S.~Sarangi and S.~H.~H.~Tye,
``Cosmic string production towards the end of brane inflation,''
Phys.\ Lett.\ B {\bf 536}, 185 (2002)
[arXiv:hep-th/0204074].

\bibitem{CMP}
E.~J.~Copeland, R.~C.~Myers and J.~Polchinski,
``Cosmic F- and D-strings,''
JHEP {\bf 0406}, 013 (2004)
[arXiv:hep-th/0312067].

\bibitem{DV}
T.~Damour and A.~Vilenkin,
``Gravitational radiation from cosmic (super)strings: Bursts, stochastic
background, and observational windows,''
arXiv:hep-th/0410222.

\bibitem{CFM}
J.~M.~Cline, H.~Firouzjahi and P.~Martineau,
``Reheating from tachyon condensation,''
JHEP {\bf 0211}, 041 (2002)
[arXiv:hep-th/0207156].

\bibitem{FH}
H.~Firouzjahi and S.~H.~Tye,
``Brane Inflation and Cosmic String Tension in Superstring Theory,''
arXiv:hep-th/0501099.

\bibitem{BEMMM}
C.~P.~Burgess, R.~Easther, A.~Mazumdar, D.~F.~Mota and T.~Multamaki,
``Multiple Inflation, Cosmic String Networks and the String Landscape,''
arXiv:hep-th/0501125.


\end{thebibliography}
\end{document}